\documentclass{emulateapj}

\usepackage{apjfonts}
\DeclareGraphicsExtensions{.pdf,.png,.jpg}

\begin{document}

\title{TRANSIT PROBABILITIES FOR STARS WITH STELLAR INCLINATION CONSTRAINTS}

\author{Thomas G. Beatty\altaffilmark{1} and Sara Seager\altaffilmark{2,3}}

\altaffiltext{1}{Department of Astronomy, The Ohio State University, 140 West 18th Avenue, Columbus, OH 43210, USA; tbeatty@astronomy.ohio-state.edu.}
\altaffiltext{2}{Department of Physics, Massachusetts Institute of Technology, Cambridge, MA 02139, USA.}
\altaffiltext{3}{Department of Earth, Atmospheric, and Planetary Sciences, Massachusetts Institute of Technology, Cambridge, MA 02139, USA.}

\slugcomment{Accepted version for publication in the Astrophysical Journal}
\shorttitle{Transit Probabilities With Stellar Inclination Constraints}
\shortauthors{Beatty \& Seager}

\begin{abstract}
The probability that an exoplanet transits its host star is high for planets in close orbits, but drops off rapidly for increasing semimajor axes. This makes transit surveys for planets with large semimajor axes orbiting bright stars impractical, since one would need to continuously observe hundreds of stars that are spread out over the entire sky. One way to make such a survey tractable is to constrain the inclination of the stellar rotation axes in advance, and thereby enhance the transit probabilities. We derive transit probabilities for stars with stellar inclination constraints, considering a reasonable range of planetary system inclinations. We find that stellar inclination constraints can improve the transit probability by almost an order of magnitude for habitable-zone planets. When applied to an ensemble of stars, such constraints dramatically lower the number of stars that need to be observed in a targeted transit survey. We also consider multiplanet systems where only one planet has an identified transit, and derive the transit probabilities for the second planet assuming a range of mutual planetary inclinations.
\end{abstract}


\section{Introduction}	

Transiting exoplanets around the brightest stars in the sky will be a major provider of science in the coming years. As the focus of the field moves toward the characterization of transiting planets, exoplanets orbiting bright stars will provide the best opportunities for precise follow-up observations. This is particularly true for any Earth-analogs discovered around the brightest Sun-like stars. The Kepler mission is surveying relatively dim stars (V = 9 to V = 16) for Earth-analogs, but these will be difficult to observe from the ground. Detecting a transiting Earth-analog around a bright (V $\leq7$) Sun-like star would enable a multitude of observations, from the ground and space, that would otherwise not be possible. Transiting planets around the brightest Sun-like stars are therefore of prime scientific interest. 

The brightest Sun-like stars have, nevertheless, yet to be comprehensively surveyed for transiting planets, Earth-analogs or otherwise. The reason is that they are spread out over the entire sky, which makes a practical survey of these stars very difficult. Since they are so widely distributed, any survey of the brightest Sun-like stars would have to be a targeted survey that examined one star at a time. Such a survey would be similar in concept to the MEarth survey of M-dwarfs \citep{nutzman2008}, the N2K radial velocity survey \citep{fishcer2005}, or the TERMS targeted transit survey \citep{kane2009}. Unfortunately, given the hundreds of stars that would need to be surveyed, the average transits probabilities, and the typical window functions for long period Earth-analogs, a targeted survey searching for long period transiting exoplanets would require either hundreds of telescopes or hundreds of years of observing time.

In the course of considering a new space-based transit search for Earth-analogs, this problem motivated us to consider ways in which a targeted survey of the brightest stars could be feasibly executed. One solution is to determine the inclination of the rotation axes of the target stars. From angular momentum considerations, we would expect any orbiting planets to lie close to the stellar equator. By only targeting those stars with stellar inclinations near $90^\circ$ to our line of sight, the number of stars that need to be surveyed could be lowered dramatically.

We have therefore calculated the effect of stellar inclination constraints on transit probabilities. We account for a range of possible planetary system inclinations with respect to the stellar equator, and apply these probabilities to an ensemble of stars. We also examine the number of stars that need to be observed to reach various statistical confidence levels of detection. For reasonable assumptions, this allows for a 85\% reduction in the number of stars that would need to be observed in a targeted survey. Furthermore, we consider the case of multiplanet systems where only one planet has a detected transit, and derive the transit probability of the second planet for various spreads of the mutual inclination angle. We conclude by discussing ways in which to measure stellar inclinations, followed by the outline of a potential space-based survey.

\section{Transit Probabilities}

\subsection{Background}

We begin by reviewing the transit probability for a single star under the assumption that the planet's orbital inclination is randomly and evenly distributed over all possible orientations. We will assume for now that the planetary orbit is coplanar with the stellar equator. In this case, the transit probability for planets in circular orbits is $R_*/a$ \citep{borucki1984}, the ratio of the stellar radius to the semimajor axis of the planetary orbit. The a priori transit probability of $R_*/a$ may be derived by considering the angular momentum vector of a planetary orbit in particular orbital orientation, the complete set of which describe a sphere in space. The probability of seeing any particular set of orientations is the fractional area on the sphere that encompasses that set's angular momentum vectors. Again, if we assume for the moment that the planetary orbit is coplanar with the stellar equator, the orbital inclination $i$ is the same as the stellar inclination $\psi$ (Figure 1). The probability of star having a particular value for its stellar inclination then follows the distribution
\begin{equation}\label{eq:10}
f_\Psi(\psi)=\sin(\psi).
\end{equation}
Transits will occur if the planetary orbit has an inclination between $(90^\circ - \theta) \leq i \leq (90^\circ + \theta)$, where the angle $\theta$ is equal to 
\begin{equation}\label{eq:20}
\theta = \arcsin \left(\frac{R_*}{a}\right),
\end{equation}
and is the maximum orbital inclination for which a planet will show a transit. Note that we have assumed the planet radius to be much smaller than the stellar radius.

\begin{figure}
\vskip -0.0in 
\epsscale{1.2} 
\plotone{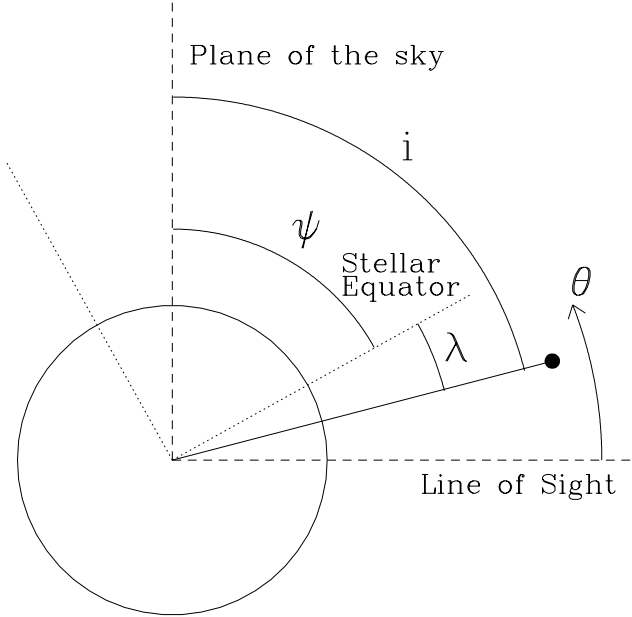}
\vskip -0.0in 
\figcaption[Relevant angles]{Diagram of the relevant angles. $\psi$ is the angle that the stellar equator makes with the plane of the sky, $\lambda$ is the angle the planetary orbit makes with the stellar equator, and $i$ is the observed orbital inclination of the planetary orbit from Earth.}
\end{figure}

The transit probability for randomly distributed circular orbits is the fraction of orbits that lie within this orbital inclination range, and is given by
\begin{equation}\label{eq:30}
P_{\mathrm{tr}} = \frac{\int_{90^\circ - \theta}^{90^\circ + \theta} \sin(\psi)\ d\psi}{\int_{0^\circ}^{180^\circ} \sin(\psi)\ d\psi} = \cos(90^\circ-\theta) = \frac{R_*}{a},
\end{equation}
which is the result quoted earlier. Note that in these and the following calculations we will continue to assume that the radius of the planet is much smaller than the radius of the star. 

\subsection{Transit Probabilities with Stellar Inclination Constraints}

When there is information about the orientation of the planetary system through measurement or indirect assumptions, the transit probability can be calculated by treating the above distribution for $\psi$ as a Bayesian prior \citep{bayes1763,cox1946}. This allows us efficiently to slot in stellar inclination measurements to our knowledge of a system's orientation. Indeed, the calculation of transit probabilities is almost a perfect subject for Bayesian techniques, since we have a rigorously defined prior distribution --- a luxury that many problems do not have.

Assuming that the planetary orbit is coplanar to the stellar equator, the probability distribution arising from a measured stellar inclination angle, $\psi_m$, can be treated as a conditional distribution. The resulting posterior distribution will be
\begin{equation}\label{eq:35}
f_\Psi(\psi|\psi_m) = \frac{\sin(\psi) f(\psi_m|\psi)}{\int_{0^\circ}^{180^\circ} \sin(\psi) f(\psi_m|\psi)\ d\psi}.
\end{equation} 
The transit probability will be the integral of the posterior distribution over the range of transiting stellar inclinations,
\begin{equation}\label{eq:40}
P_{\mathrm{tr}} = \int_{90^\circ - \theta}^{90^\circ + \theta} f_\Psi(\psi|\psi_m)\ d\psi.
\end{equation} 

As an example, take a Gaussian measurement of the form $\psi_m \pm \sigma$. This implies a conditional distribution of
\begin{equation}\label{eq:50}
f_\mathrm{Gauss}(\psi_m|\psi) = \frac{1}{\sqrt{2 \pi \sigma^2}} \exp \left[\frac{-(\psi - \psi_m)^2}{2\sigma^2}\right],
\end{equation}
and a posterior of
\begin{equation}\label{eq:60}
f_\mathrm{Gauss}(\psi|\psi_m) = \frac{\frac{\sin(\psi)}{\sqrt{2 \pi \sigma^2}} \exp \left[\frac{-(\psi - \psi_m)^2}{2\sigma^2}\right]}{\int_{0^\circ}^{180^\circ} \frac{\sin(\psi)}{\sqrt{2 \pi \sigma^2}} \exp \left[\frac{-(\psi - \psi_m)^2}{2\sigma^2}\right]\ d\psi},
\end{equation}
If we have made an inclination measurement of $\psi_m=90^\circ \pm 5^\circ$ with Gaussian uncertainties, then for an Earth-like planet with $R_*/a = 1/215$ the transit probability will be 4.25\%. This is 9 times greater than the transit probability before the inclination measurement was made (0.47\%). For hot Jupiters with $R_*/a=1/10$, the transit probability is enhanced to 74.8\% from 10\%. 

Alternatively, consider a box-like, uniform, distribution of the measured angle. This could occur if we had a ``binary'' measurement that only revealed if the stellar inclination was above or below a certain angle. In this case, the conditional distribution from the measurement of $\psi_m$ is the combination of two Heaviside step functions
\begin{equation}\label{eq:70}
f_\mathrm{box}(\psi_m|\psi) = \frac{1}{2 \sigma} (H[\psi + (-\psi_m+ \sigma)]\ H[-\psi + (\psi_m + \sigma)]). 
\end{equation} 
The posterior will be
\begin{eqnarray}\label{eq:80}
&&f_\mathrm{box}(\psi|\psi_m) = \nonumber \\
&& \frac{\frac{\sin(\psi)}{2 \sigma} (H[\psi+ (-\psi_m+ \sigma)]\ H[-\psi + (\psi_m + \sigma)])}{\int_{0^\circ}^{180^\circ} \frac{\sin(\psi)}{2 \sigma} (H[\psi + (-\psi_m+ \sigma)]\ H[-\psi + (\psi_m + \sigma)])\ d\psi}. 
\end{eqnarray} 
For a similar measurement of $\psi_m=90^\circ \pm 5^\circ$, the transit probability for an Earth-like planet at $R_*/a = 1/215$ is 5.3\%, an increase of a factor of 11.4 over the transit probability before making the measurement. hot Jupiters, on the other hand, are close enough to the star that they will always transit across a star with a measured stellar inclination of $\psi_m=90^\circ \pm 5^\circ$.

\subsection{Transit Probabilities for a Range of Planetary System Inclinations}

We now consider the added complexity of inclination in the planetary system itself. In the preceding discussion we assumed that any orbiting planets are coplanar with the stellar equator. From the solar system we know that this is not necessarily the case. The ecliptic is inclined by $7.155^\circ$ to the Sun's equator, and the rest of the solar system's planets are scattered within several degrees of the Ecliptic. We will define the inclination of the planetary system to the stellar equator --- or any other reference plane --- as the variable $\lambda$. The angle the stellar equator makes with the plane of the sky we denote as the stellar inclination $\psi$, and we will to refer to the observed angle the planetary orbit makes with the plane of the sky as the orbital inclination $i$ (Figure 1).

Factoring in the planetary inclination of the exoplanetary system, the orbital inclination (i.e., as seen from Earth) will be $i=\psi-\lambda$, and its distribution will be governed by the joint distribution of $\psi$ and $\lambda$. Since $\psi$ and $\lambda$ are independent variables, their joint distribution is simply the product of their individual distributions: $f_J = f_{\Psi} \cdot f_\Lambda$, where $f_J$ is the joint distribution, $f_\Psi$ is the stellar inclination distribution, and $f_\Lambda$ is the planetary inclination distribution. We will substitute $\psi -i$ for $\lambda$ in the distribution $f_\Lambda$ to help the calculations. Making the substitution, the distribution of the angle $i$ is then
\begin{equation}\label{eq:620}
f_\mathrm{I}(i) = \int_{-\infty}^{\infty} f_{\Psi}(\psi|\psi_m) f_\Lambda(\psi -i) d\psi.
\end{equation}  

The function $f_I(i)$ is called the marginal probability distribution for the orbital inclination. To calculate the transit probability of a system, $f_I(i)$ becomes the integrand in Equation (\ref{eq:40}):
\begin{equation}\label{eq:630}
P_{\mathrm{tr}} = \frac{\int_{90^\circ - \theta}^{90^\circ + \theta} f_I(i)\ di}{\int_{0^\circ}^{180^\circ} f_I(i)\ di}.
\end{equation} 
Here $P_{tr}$ is the probability that a single star will show a transit, assuming that there is a planet in orbit. Equations (\ref{eq:620}) and (\ref{eq:630}) may also be used on multiple planet systems: if one planet is observed to transit the parent star, its orbital inclination may be used as a reference plane in place of the stellar inclination $\psi_m$, and an assumed mutual inclination distribution may be substituted for the planetary system inclination $\lambda$. The transit probability of a second planet in the system may then be calculated. The single-star transit probability may also be applied to an ensemble of stars with random stellar inclinations.

\subsection{Eccentric Planetary Orbits}

We have assumed up to this point that any potential planets are on circular orbits. To calculate the transit probability for eccentric planetary orbits, consider the opening angle $\theta$ which describes the maximum angular difference between our line of sight and the observed orbital inclination, $i$, that allows a planetary transit. In the case of a circular orbit $\theta$ is defined as in Equation (\ref{eq:20}). In the more general case, $\theta$ is   
\begin{equation}\label{eq:2400}
\theta = \arcsin \left(\frac{R_*}{r}\right),
\end{equation} 
where $r$ is the planet-star separation at the time of transit. For an eccentric Keplerian orbit, this implies that
\begin{equation}\label{eq:2410}
\theta = \arcsin \left(\frac{R_*}{a} \frac{1+e\sin(\omega)}{1-e^2}\right).
\end{equation}
Here $e$ is the orbital eccentricity and $\omega$ is the argument of periastron. This has been noted previously by \cite{barnes2007} and \cite{burke2008}.

\begin{figure}
\vskip -0.0in 
\epsscale{1.2} 
\plotone{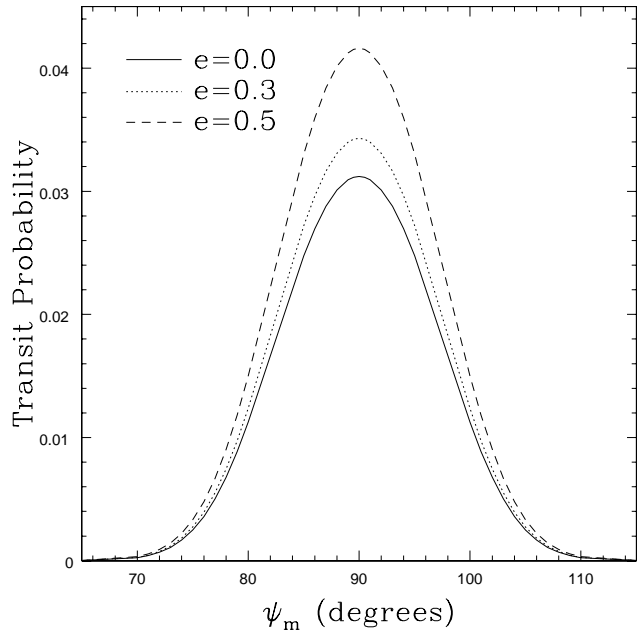}
\vskip -0.0in 
\figcaption[Eccentric orbits]{Eccentricity effects on the transit probability for an Earth-analog, as a function of the measured stellar inclination angle. Note that these are values averaged over an assumed uniform distribution for the argument of periastron. We have used a measurement uncertainty of 5$^\circ$ and planetary inclinations spread uniformly within 7.5$^\circ$ of the stellar equator.}
\end{figure}

In the context of transit probabilities with stellar inclination constraints, the effect of an eccentric planetary orbit will be to change the integration limits in Equation (\ref{eq:630}). The exact effect will depend upon the values of $e$ and $\omega$, or more generally upon the distribution of $e$ and $\omega$ in exoplanetary systems. In the situations we are interested in, when there is no radial velocity evidence for a planet, we expect the distribution for the argument of periastron, $\omega$, to be evenly distributed. This implies that on average any orbital eccentricity will act as a ``boost factor'' to the angle $\theta$ of $1/(1-e^2)$.  

Figure 2 demonstrates how this boost factor affects the transit probability of an Earth-analog, as a function of the measured stellar inclination angle. We have averaged over an assumed uniform distribution for the argument of periastron, and used a measurement uncertainty of $5^\circ$. The planets are also assumed to have planetary inclinations evenly spread within $7.5^\circ$ of the stellar equator. At $e=0.3$, the boost factor to $\theta$ (and, to a very near approximation, the transit probability) is 1.10. An eccentricity of 0.5 gives a boost of 1.33. 

Note that for hot Jupiters and Earth-analogs --- the two cases we treat subsequently --- we expect the planets to be in nearly circular orbits, with $e\le0.1$. An eccentricity of $e=0.1$ provides a boost factor of 1.01 to the angle $\theta$. Fifty out of the 57 transiting hot Jupiters with measured eccentricities have $e\le0.1$, and although the eccentricity distribution of Earth-analogs is poorly constrained and outside the scope of this paper, we note that Earth has an eccentricity of $e=0.017$. We therefore only consider circular orbits in Section 3.2, when we calculate the benefits stellar inclination constraints provide in transit searches for hot Jupiters and Earth-analogs. 

\subsection{Transit Probabilities for an Ensemble of Stars}

One of our goals is to determine the number of stars one must survey to find a transiting planet, given a fixed uncertainty in the measured stellar inclinations and an adopted range of planetary inclinations. To do this, we first need to determine the size of our initial target list, $n_\mathrm{i}$, for stellar inclination measurements. This will set the probability $P_\mathrm{ntr}$ that there is the desired number of transiting planets in the stars that we are considering. For instance, if we measured the stellar inclination of just 10 stars there would be a low probability that a transiting Earth-analog is in our sample, but measuring the stellar inclination of several thousand stars would make it near certain.

Second, we also want to calculate the fraction of the initial target stars that will actually need to be observed photometrically for transits. This fraction will set the odds that we will observe one of the transiting planets hidden within our original set of stars, which we denote as $P_\mathrm{obs}$. Since we are truly interested in only those stars with stellar inclinations near $90^\circ$, the size of this fraction will depend directly on the precision of the stellar inclination measurements. It will also be sensitive to assumptions made about the distribution of $\lambda$, the planetary system inclination. Later, it will be useful to delineate the fraction of the initial stars that we will observe photometrically by the angle $\phi$. We then will observe all of the stars with measured stellar inclinations within $\phi$ degrees of $90^\circ$. 

We first turn to describing $P_\mathrm{ntr}$, the probability that there is the desired number of transiting planets within our initial $n$ target stars. We will calculate $P_\mathrm{ntr}$ using the binomial distribution. Previous studies have instead used the expectation value to determine the number of stars that one needs for a transit detection: if hot Jupiters have a 10\% transit probability, then a survey will need to look at ten stars to statistically expect a single detection. We use the binomial distribution to perform a similar calculation. Unlike using the expectation value, the binomial distribution allows us to explicitly place a probability that the survey will detect the desired number of transiting planets. This calculation will be based upon the a priori transit probability denoted as $P_{tr,0} = R_*/a$, since we are trying to determine how many stars we need to start with in our initial sample to yield enough transiting planets. We will denote the binomial distribution for $s$ successes in $n$ independent trails, each with success probability $p$, as the function
\begin{equation}\label{eq:2510}
\mathrm{Bi}(n,s,p) = \left(\begin{array}{c} n \\ s \end{array}\right) \left(p\right)^s \left(1-p\right)^{n-s}.
\end{equation}

If we start with $n_\mathrm{i}$ initial target stars, consider the probability $P_\mathrm{det,1}$ that there is exactly one transiting planet in the initial stars, and that we detect it. This probability will be the product of the odds $P_\mathrm{ntr,1}$ that there is exactly one transiting planet in the $n_\mathrm{i}$ initial stars, and the odds that we will observe it photometrically, $P_\mathrm{obs}$. Using the binomial distribution to calculate the chance that there is exactly one transiting planet in our initial stars, we can write    
\begin{equation}\label{eq:2520}
P_\mathrm{det,1} = P_\mathrm{ntr,1}\ P_\mathrm{obs} = \mathrm{Bi}(n_\mathrm{i},1,P_\mathrm{tr,0})\ P_\mathrm{obs}.
\end{equation}  

Now let us take the case of exactly three transiting planets among our initial stars. What is the probability that we will detect at least one of them? Again we may use the binomial distribution to calculate the probability $P_\mathrm{ntr,3}$ that there are exactly three transiting planets, but now we will also use a sum of the binomial distribution to determine the probability that we observe at least one of these planets. We take the sum because we will regard observing one, two, or all three of the planets as successfully fulfilling our criteria that we observe at least one planet. We therefore add the probabilities for all three of these scenarios, and
\begin{equation}\label{eq:2530}
P_\mathrm{det,3} = \mathrm{Bi}(n_\mathrm{i},3,P_\mathrm{tr,0})\ \sum_{k=1}^3\mathrm{Bi}(3,k,P_\mathrm{obs}).
\end{equation} 

The general case, and the one that we are ultimately interested in, is the one wherein we wish to know the probability that we will detect at least $n_\mathrm{tr}$ transiting planets, but do not specify the exact number of transiting planets in our initial $n_\mathrm{i}$ stars. To calculate this overall probability we must sum the individual probabilities that we will detect at least $n_\mathrm{tr}$ planets given an exact number of transiting planets in our $n_\mathrm{i}$ stars. This similar to how we summed the probability of observation in Equation (\ref{eq:2530}). In terms of our initial target stars, the probability of photometrically observing one transiting planet, and the number of transiting planets that we wish to detect, the probability that we will detect at least that many planets is therefore   
\begin{equation}\label{eq:2540}
P_\mathrm{det}(n_\mathrm{i},P_\mathrm{obs},n_\mathrm{tr}) = \sum_{j=n_\mathrm{tr}}^{n_\mathrm{i}} \left(\mathrm{Bi}(n_\mathrm{i},j,P_\mathrm{tr,0})\ \sum_{k=n_\mathrm{tr}}^j\mathrm{Bi}(j,k,P_\mathrm{obs})\right).
\end{equation}  

We are still left with the calculation of $P_\mathrm{obs}$, the independent probability that we will observe each of the transiting planets within our initial $n_\mathrm{i}$ stars. We will only photometrically observe a fraction of the initial $n_\mathrm{i}$ target stars for transits. These will be the stars with measured inclinations within $\phi$ degrees of $90^\circ$, and the size of the fraction will be determined by the desired level of $P_{obs}$. A survey will only photometrically observe stars with measured stellar inclinations between $90^\circ-\phi \leq \psi_m \leq 90^\circ+\phi$. Now consider that within the total number of stars, $n_\mathrm{i}$, that will have their stellar inclinations measured we expect that the measured inclinations will be randomly distributed around the sky, and so the distribution of $\psi_m$ will go as $\sin(\psi_m)$. The expected number of stars that will lie within $\phi$ degrees of $90^\circ$ will therefore be
\begin{equation}\label{eq:5400}
n_\mathrm{obs} = n_\mathrm{i} \left(\frac{\int_{90^\circ-\phi}^{90^\circ+\phi} \sin(\psi_m)\ d\psi_m}{\int_{0}^{180} \sin(\psi_m)\ d\psi_m}\right) = n_\mathrm{i} \sin(\phi).
\end{equation}
To find the number of stars, $n_{obs}$ that will need to be observed for transits, we will need to calculate $n_\mathrm{i}$ and $\phi$. Note that we are considering the general case and therefore leave out survey-specific considerations that should be included for a more accurate yield estimate. For example, we will assume that every star has a planet of the type we are looking for, whereas the true frequency will depend upon the period or mass range that the survey is most sensitive to. We also do not consider the effect of non-central transits or variations in stellar-type, since these will both be directly tied into the signal-to-noise and magnitude limits of a given survey. For a more detailed discussion on how these factors can affect the yields of transit surveys, see \cite{beatty2008}.

To calculate $\phi$, the range of measured stellar inclinations that will be considered for transit observations, first consider the transit probability as a function of the measured stellar inclination of the target star. This will allow us to calculate the probable distribution of the transiting planets as a function of the measured stellar inclination, and thereby figure what stellar inclination range will have to observed to recover a given fraction of the transiting planets within our sample of $n_\mathrm{i}$ stars.

Including the inclination of the planetary systems as per Equation (\ref{eq:620}), the transit probability will be
\begin{equation}\label{eq:5100}
P_{\mathrm{tr}}(\psi_m) = \frac{\int_{90^\circ-\theta}^{90^\circ+\theta} \int_{0^\circ}^{180^\circ} f_{Gauss}(\psi|\psi_m) f_\Lambda(\psi-i)\ d\psi \ di}{\int_{0^\circ}^{180^\circ} \int_{0^\circ}^{180^\circ} f_{Gauss}(\psi|\psi_m) f_\Lambda(\psi-i)\ d\psi \ di},
\end{equation}
where $f_{Gauss}(\psi|\psi_m)$ is the posterior distribution of the stellar inclination. We now multiply $P_{\mathrm{tr}}$ by the probability that measuring a star yields a given stellar inclination angle, $\sin(\psi_m)$. Doing so will give us the probability that measuring the stellar inclination of a star yields a certain value, and that with a measured stellar inclination $\psi_m$ the star will then show a transiting planet,
\begin{equation}\label{eq:5200}
P_{\Psi_m}(\psi_m) = \sin(\psi_m) P_{\mathrm{tr}}(\psi_m).
\end{equation} 
The distribution $P_{\Psi_m}$ is, physically, the expected number of planets --- normalized to unity --- that we will expect to see transiting stars with a given measured stellar inclination $\psi_m$.

The fraction of transits that we will recover is dependent upon how much of this distribution is observed for transits. If we observe all of the stars with measured stellar inclinations within an angle $\phi$ of $90^\circ$, then
\begin{equation}\label{eq:5300}
P_\mathrm{obs} = \frac{\int_{90-\phi}^{90+\phi} \sin(\psi_m)\ P_{\mathrm{tr}}(\psi_m)\ d\psi_m}{\int_{0}^{180} \sin(\psi_m)\ P_{\mathrm{tr}}(\psi_m)\ d\psi_m}.
\end{equation}  
For a given value of $P_\mathrm{obs}$, we may solve for the angle $\phi$ that defines the outer limit of the subsample observed for transits.

\begin{figure}
\vskip -0.0in 
\epsscale{1.2} 
\plotone{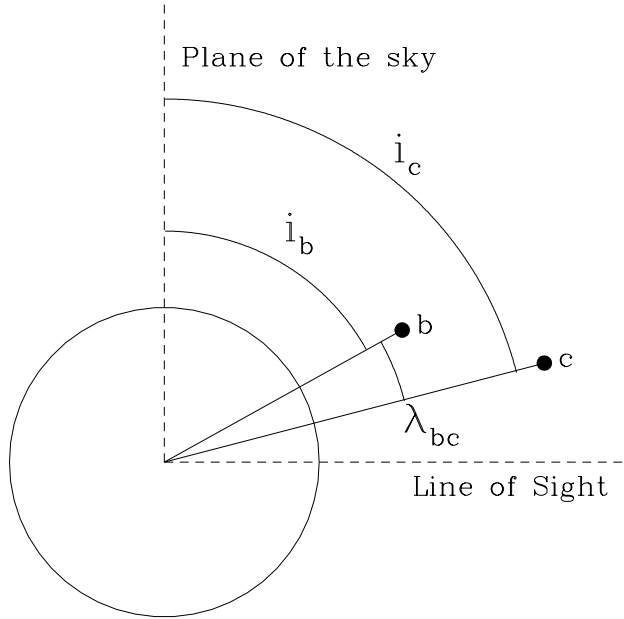}
\vskip -0.0in 
\figcaption[Multiple Planet System]{Diagram of a notional multiplanet system displaying the angles referenced in the text. The angles $i_\mathrm{b}$ and $i_\mathrm{c}$ are the orbital inclinations of the planets as observed from Earth. $\lambda_{\mathrm{bc}}$ is the mutual inclination angle between the two orbits.}
\end{figure}

We can therefore calculate $n_\mathrm{i}$ and $\phi$ for a given survey by specifying the desired number of transiting planet detections, and the desired values of $P_\mathrm{obs}$. This allows us to determine $n_\mathrm{obs}$, the number of planets that will have to be observed for transits, as well as the probability that we will detect the desired number of transiting planets, 
\begin{equation}\label{eq:5320}
P_\mathrm{det}(n_\mathrm{i},P_\mathrm{obs},n_\mathrm{tr}) = \sum_{j=n_\mathrm{tr}}^{n_\mathrm{i}} \left(\mathrm{Bi}(n_\mathrm{i},j,P_\mathrm{tr,0})\ \sum_{k=n_\mathrm{tr}}^j\mathrm{Bi}(j,k,P_\mathrm{obs})\right)
\end{equation} 

\section{Results}

\subsection{Probability for a Second Planet to Transit} 

One direct application of our probability calculations is in determining the transit probability of a second planet in a system in which one planet is already observed to transit the parent star. From angular momentum considerations, we have been using the stellar equatorial plane as the reference plane for our assumed distributions for the inclination of the planetary systems, $f_\Lambda(\lambda)$. In the case of a multiple planet system we may instead use the plane of the transiting planet's orbit as the reference. Consider the planetary system shown in Figure 3. We may measure the orbital inclination $i_{b,m}$ of the transiting inner planet through the transit photometry. Assuming Gaussian uncertainties, this will give us a distribution $f_{I_\mathrm{b}}(i_\mathrm{b}|i_{\mathrm{b,m}})$ for the orbital inclination of the inner planet. The outer planet will be inclined at some angle $\lambda_{\mathrm{bc}}$ to the inner planet, and will have an orbital inclination with respect to the sky of $i_\mathrm{c} = i_\mathrm{b} + \lambda_{\mathrm{bc}}$. The mutual inclination of the two planets $\lambda_{\mathrm{bc}}$ will in turn be drawn from the distribution $f_{\Lambda_{\mathrm{bc}}}(\lambda_{\mathrm{bc}})$.

\begin{figure}
\vskip -0.0in 
\epsscale{1.2} 
\plotone{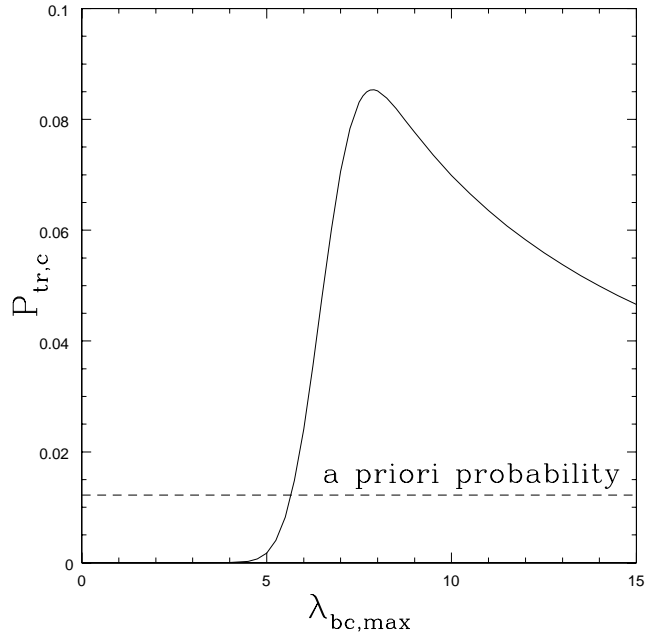}
\vskip -0.0in 
\figcaption[HAT-P-13c Transit Probability]{Transit probability of HAT-P-13c, as a function of the maximum mutual inclination of the two planets ($\lambda_{\mathrm{bc,max}}$). This assumes that the orbit of planet `c' is evenly distributed within $\lambda_{\mathrm{bc,max}}$ degrees of planet`b's orbit. For reference, the a priori transit probability of $R_*/r_\mathrm{c}$ is also plotted.}
\end{figure}

Inserting these distributions into Equation (\ref{eq:620}) gives the marginal distribution of the orbital inclination for the outer planet (here we replace $\psi_m$ with the angle $i_\mathrm{b}$ as the relevant reference plane for the system)
\begin{equation}\label{eq:4100}
f_{I_\mathrm{c}}(i_\mathrm{c}) = \int_{-\infty}^{\infty} f_{I_\mathrm{b}}(i_\mathrm{b}|i_{b,m}) f_{\Lambda_{\mathrm{bc}}}(i_\mathrm{c}-i_\mathrm{b})\ di_\mathrm{b}.
\end{equation}  
The transit probability will be 
\begin{equation}\label{eq:4200}
P_{\mathrm{tr}} = \frac{\int_{90^\circ-\theta_\mathrm{c}}^{90^\circ+\theta_\mathrm{c}} f_{I_\mathrm{c}}(i_\mathrm{c}) \ di_\mathrm{c}}{\int_{0^\circ}^{180^\circ} f_{I_\mathrm{c}}(i_\mathrm{c}) \ di_\mathrm{c}}
\end{equation}
as per Equation (\ref{eq:630}). The angle $\theta_\mathrm{c}$ is determined by the orbital separation of planet `c' and the host star at the time of their conjunction, $\theta_\mathrm{c} = \arcsin(R_*/r_\mathrm{c})$.

As an example, consider the HAT-P-13 system \citep{bakos2009}. As of this writing\footnote{2009 September} only the inner planet, HAT-P-13b, has been observed to transit. What is the probability that the outer planet. HAT-P-13c, also transits? \cite{bakos2009} measure the orbital inclination of the inner planet to be $i_{\mathrm{b,m}} = 83.4^\circ \pm 0.6^\circ$. At the time of its conjunction with HAT-P-13, the outer planet is at a distance of $r_\mathrm{c}/R_* = 82.1 \pm 6.1$ stellar radii from the star. This allows us to determine $f_{I_\mathrm{b}}(i_\mathrm{b}|i_{\mathrm{b,m}})$ and $\theta_\mathrm{c}$ for the HAT-P-13 system. For the mutual inclination of the two planets, we assumed that the orbital inclination of planet `c' was evenly distributed within $\pm \lambda_{\mathrm{bc,max}}$ degrees of the orbital inclination $i_\mathrm{b}$ of planet `b'. Figure 4 shows the transit probability of HAT-P-13c as a function of $\lambda_{\mathrm{bc,max}}$. 

The probability that planet `c' transits is very dependent upon what we assume is a reasonable range of mutual inclination in the HAT-P-13 system. For reference, all of the Solar System planets are within $3.4^\circ$ of the Earth's orbit --- except for Mercury (at $7^\circ$). If the mutual inclination of the two planets orbiting HAT-P-13 is within $3.4^\circ$, then the outer planet will not transit. If the two planets are misaligned by up to $7^\circ$, then the transit probability for planet `c' is 7\%. The maximum transit probability of 8.5\% occurs if we assume that the two planets may be inclined within $8^\circ$ of each other. As the assumed spread in mutual inclination increases, the transit probability will fall back to the a priori value of 1.2\%. 

The transit probability for HAT-P-13c is therefore at most 8.5\%.

\subsection{Benefits for Target Selection}

We now demonstrate how using stellar inclination measurements and the enhanced transit probabilities (Section 2) can aid in the target selection of transit surveys. As illustrative cases, we will calculate how many stars need to be observed in a survey looking for hot Jupiters, and for a separate survey searching for planets within the habitable-zone. In these examples, we will make the simplifying assumption that every star has either a hot Jupiter or habitable-zone planet in orbit, at distances of $R_*/a=1/10$ or $R_*/a=1/215$, respectively. 

We assume that the inclinations of the planetary systems are distributed in two ways. For the hot Jupiters, we use the planetary inclination distribution determined by \cite{fabrycky2009} from an ensemble of 11 Rossiter---McLaughlin measurements of spin---orbit alignment. The authors found that aside from the XO-3 system\footnote{Which appears to be highly misaligned, possibly because of XO-3b's migration history.} the hot Jupiters they considered had planetary inclinations distributed according to a Rayleigh distribution with a width parameter of $6.6^\circ$. Exoplanets within the habitable-zone may not follow this same planetary inclination distribution. We use a uniform distribution of planetary inclination within $7.5^\circ$ of the stellar equator; Earth has an planetary inclination of $7.155^\circ$ to the Sun's equator. Including either spread of planetary inclinations into the calculations of transit probabilities acts to spread out the probability of transit, and make stars with stellar inclinations far from $90^\circ$ more likely to show transits. At the same time, the spread of planetary inclinations makes stars with measured stellar inclinations near $90^\circ$ less likely to show transits. Assuming a measurement of $\psi_m=90^\circ \pm 5^\circ$, and that planetary inclinations are uniformly distributed within $7.5^\circ$ of the stellar equator, then the transit probability for a habitable-zone planet at a distance of $R_*/a = 1/215$ star drops from 4.25\% to 3.08\% as compared to assuming the orbit is coplanar with the stellar equator. Conversely, a star with measured stellar inclination of $\psi_m=80^\circ \pm 5^\circ$ has its transit probability increased from 0.59\% to 1.11\% 

As our first example, take a survey for hot Jupiters around solar-type stars. We will require a 95\% probability that the survey detects at least a single transiting planet. From Equation (\ref{eq:2540}) this means that $P_\mathrm{det}=0.95$. Although we may arbitrarily set $P_\mathrm{obs}$ and $n_\mathrm{i}$, as shown in the appendix the time required to complete a hot Jupiter survey is minimized if we set $P_\mathrm{obs}=0.9814$. To find the number of stars needed in the initial sample we must then solve
\begin{equation}\label{eq:5600}
P_\mathrm{det} = 0.95 = \sum_{j=1}^{n_\mathrm{i}} \left(\mathrm{Bi}(n_\mathrm{i},j,\frac{1}{10})\ \sum_{k=1}^j\mathrm{Bi}(j,k,0.9814)\right).
\end{equation}  
We must therefore have $n_\mathrm{i}=29$ stars in our initial sample.

We next want to know how many stars out of these 29 will actually have to be observed photometrically. That is, how many of the initial targets with measured inclinations near $90^\circ$ will we need to look at for transits? We will assume that all of the stellar inclination measurements have Gaussian uncertainties of $5^\circ$. The transit probability for a hot Jupiter can be calculated for various orientation measurements of the form $\psi_m \pm 5^\circ$ by using Equation (\ref{eq:620}) to account for the inclination of the planetary system:
\begin{equation}\label{eq:5700}
P_{\mathrm{tr}}(\psi_m) = \frac{\int_{90^\circ-5.73^\circ}^{90^\circ+5.73^\circ} \int_{0^\circ}^{180^\circ} f_{Gauss}(\psi|\psi_m) f_\Lambda(\psi -i)\ d\psi \ di}{\int_{0^\circ}^{180^\circ} \int_{0^\circ}^{180^\circ} f_{Gauss}(\psi|\psi_m) f_\Lambda(\psi-i)\ d\psi \ di}.
\end{equation} 
hot Jupiters at a distance of $R_*/a = 1/10$ will show transits up to a maximum angle of $\theta = 5.73^\circ$. The angle $\phi$ that defines our observed subsample solves
\begin{equation}\label{eq:5800}
P_{obs} = 0.9814 = \frac{\int_{90-\phi}^{90+\phi} \sin(\psi_m)\ P_{\mathrm{tr}}(\psi_m)\ d\psi_m}{\int_{0}^{180} \sin(\psi_m)\ P_{\mathrm{tr}}(\psi_m)\ d\psi_m},
\end{equation} 
and is $\phi=24.01^\circ$ 

To detect at least one hot Jupiter, we must therefore photometrically observe
\begin{equation}\label{eq:5900}
n_\mathrm{obs} = 29 \sin(24.01^\circ) = 11.80
\end{equation}
stars that will have measured stellar inclinations within $24.01^\circ$ of $90^\circ$. This will give us a probability of $P_\mathrm{det}=95$\% of detecting at least one hot Jupiter. The top panel of Figure 5 shows how the number of stars that need to be observed varies as a function of the stellar inclination measurement precision for various confidence levels. The top panel also shows how the fraction of the initial target list that will need to be observed varies with measurement precision. The lower limits in both cases set by the spread in the distribution of $f_\Lambda(\lambda)$ as determined by \cite{fabrycky2009}.

In a notional survey for habitable-zone planets, we will also require a 95\% chance of detecting at least one transit. The calculations are similar to those for the survey for hot Jupiters, except that we will assume that $f_\Lambda(\lambda)$, the distribution of planetary system inclinations, is evenly distributed within $7.5^\circ$ of the stellar equator. Otherwise, we keep the requirement that $P_\mathrm{det}=0.95$, and optimize $P_\mathrm{obs}$ in the same way. The time needed to complete a survey for Earth analogs is minimized if we set $P_{obs}=0.7746$, which means the initial sample size will be $n_\mathrm{i}=830$ stars. Proceeding in the same way as the hot Jupiter survey, we will need to observe stars with measured stellar inclinations out to $\phi=8.12^\circ$ --- corresponding to $n_\mathrm{obs}=117.2$ stars that will need to be observed photometrically. The lower two panels of Figure 5 show how $n_\mathrm{obs}$ and how the fraction of initial stars that need to be observed change as a function of measurement precision, assuming the planetary inclinations are within $7.5^\circ$ of the stellar equator.

\begin{figure}
\vskip -0.0in 
\epsscale{1.2} 
\plotone{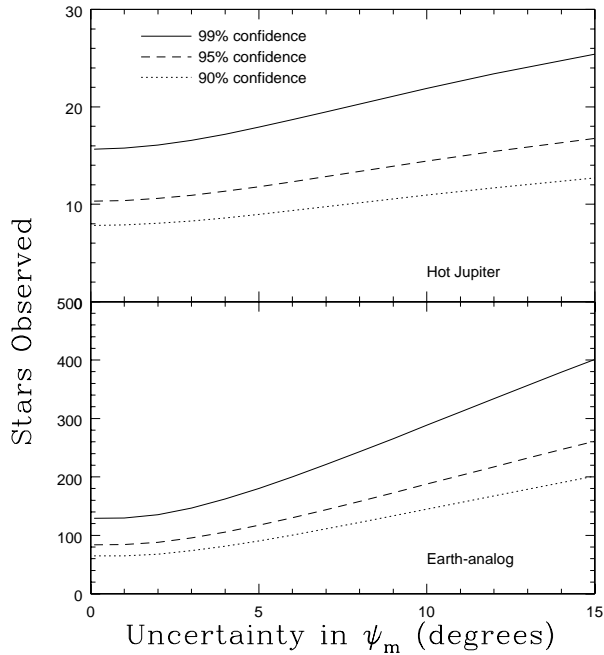}
\vskip -0.0in 
\figcaption{Number of stars that would need to be photometrically observed in a survey for different detection confidence levels. Note that this assumes every star has a corresponding planet.}
\end{figure}

Unlike the case of the hot Jupiters, there is no statistically compelling information regarding the inclination distribution, $f_\Lambda$, of Earth-analogs. Since the fraction of the stars that would need to be observed photometrically is tied to this distribution, Figure 6 illustrates the effect of changing the maximum spread of the planetary inclination on the curves from the bottom panel of Figure 5. In the above calculations we used a uniform distribution for $f_\Lambda$ out to a maximum of $7.5^\circ$, but here we also plot the fraction of stars that will need to observed against stellar inclination measurement precision for a maximum of $15^\circ$ and $0^\circ$ (i.e. perfectly aligned). Note the difference between the three curves becomes most pronounced with higher precision (lower uncertainty) inclination measurements as the underlying distribution for $f_\Lambda$ becomes more dominant. Under our assumption that the uncertainty in $\psi_m$ is $\pm 5^\circ$, the fraction of stars that need to be observed is 0.11, 0.14 and 0.21 for maximum inclinations of $0^\circ$, $7.5^\circ$ and $15^\circ$, respectively.

\begin{figure}
\vskip -0.1in 
\epsscale{1.25} 
\plotone{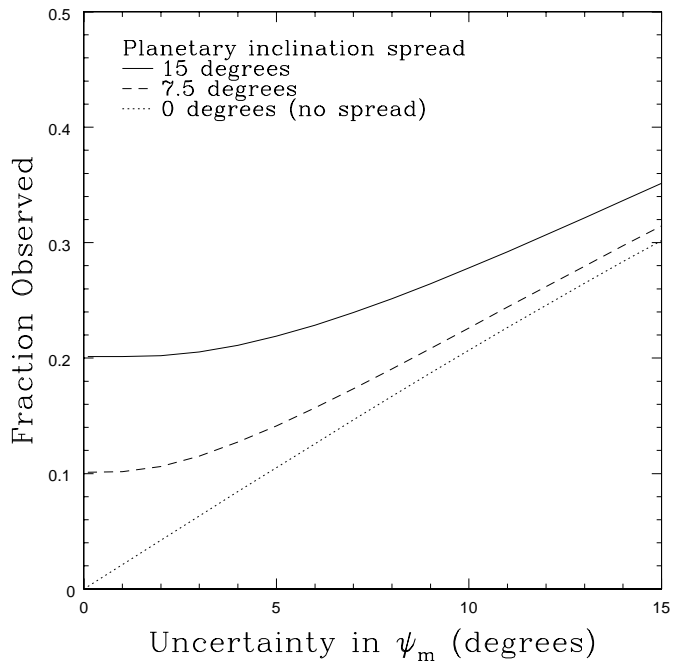}
\vskip -0.0in 
\figcaption{Fraction of stars from the initial target list that need to observed photometrically as a function of the uncertainty in measuring the stellar inclination. The three curves plotted correspond to differing maximums for the spread of the planetary inclination distribution.}
\end{figure}

In both of our example surveys, the number of stars that need to be observed photometrically is dramatically lower than the initial sample size. The exact difference will depend upon the precision of the stellar inclination measurements; we have assumed Gaussian uncertainties of $5^\circ$. The uncertainties themselves depend upon the method used to measure stellar inclinations. The two most prominent ways to measure stellar inclinations are either through spectroscopic measurements of stellar $v\sin(\psi)$ or through asteroseismic observations.

\section{Discussion}

\subsection{Spectroscopic $v\sin(\psi)$ Inclination Measurements}

Spectroscopic measurements of $v\sin(\psi)$ allow us to measure stellar inclinations indirectly. If the true rotational period of the star can be identified through photometric variation --- or other means --- and we have an estimate for the radius of the star, we will be able to calculate the value of $\sin(\psi)$. Recently \cite{winn2007} and \cite{arentoft2008} have measured $\sin(\psi)$ for two different stars. The Arentoft group's work is also particularly illustrative of the potential pitfalls in any attempt to measure $\sin(\psi)$: they observed photometric variation on Procyon with a period of $10.3\pm0.5$ days, but found it much more likely that the true rotational period of Procyon is twice that value. 

The use of $v\sin(\psi)$ measurements to constrain stellar inclinations is limited by the uncertainties in the measurement of $v\sin(\psi)$, the stellar radius, and the stellar rotational period. Spectroscopic measurements of $v\sin(\psi)$ for solar-like stars typically have fractional uncertainties of 15-25\% \citep{keppens1995,terndrup2002}. The precision may be increased by using extremely high resolution spectra ($R\approx$100,000; Carney et al. 2008) to disentangle the line variations caused by rotation and turbulence. To measure the rotational period of a star, one would either need clearly identifiable photometric variation, or would have to observe a star for several months to identify the rotation period in a periodogram \citep{meibom2009}. Together with fractional uncertainties in the stellar radius and rotational period, it is therefore time intensive to measure $\sin(\psi)$ to better than 15-30\%. 

Using measurements of $\sin(\psi)$ to constrain stellar inclinations is complicated by the flatness of the sine function near 90$^\circ$. Consider that $\arcsin(0.9)=64^\circ$ and that $\arcsin(0.995)=84^\circ$. Any determination of $\sin(\psi)$ will therefore need to be extremely precise to yield usable constraints on angles near $\psi=90^\circ$. As noted previously, this level of precision would require specialized, time-intensive observations. However, measurements of $v\sin(\psi)$ can also be used to identify and eliminate from consideration stars with stellar inclinations far from $90^\circ$. Assuming a 15\% fractional error on $\sin(\psi)$, one would be able to identify 28\% of stars as unsuitable. 

\subsection{Asteroseismic Inclination Measurements}

Stellar inclinations may also be measured directly through precise asteroseismological measurements of solar-like oscillations. These 5 minute acoustic oscillations in the stellar photosphere can be described as spherical harmonics with the harmonic numbers $n$, $l$, and $m$. On a non-rotating star the $2l+1$ $m$-modes are degenerate and lay on top of one another in frequency-space. As the angular velocity of the star increases, the $m$-modes undergo rotational---splitting, and pull apart from one another \citep{ledoux1951}. Additionally, \cite{gizon2003} show that the relative power in each of the $m$-modes will depend upon the stellar inclination. By measuring the magnitude of the splitting and the relative power of the split modes, it is possible to determine the angular velocity and stellar inclination of a star that undergoes solar-like oscillations. 

Asteroseismic observations can be conducted using either photometry or spectroscopy. \cite{gizon2003} and \cite{ballot2008} describe the theoretical basis and the expected uncertainties in measuring stellar inclinations using photometric asteroseismological observations. Problematically for surveys of Sun-like stars, both papers calculate that stars with angular velocities near solar are extremely challenging targets. In \cite{gizon2003} the formal error on measurements of the stellar inclination can be very large as a star's angular velocity approaches that of the Sun. Nevertheless, photometric asteroseismology is being conducted from space by the CoRoT, Kepler, and MOST missions, and on the ground by numerous observers. To date, the only photometric detection of rotational splitting in a star other than the Sun has been accomplished photometrically using the CoRoT spacecraft \citep{appourchaux2008}; though the authors note inconsistencies in their data as compared to earlier observations of the same star \citep{mosser2005}.

Spectroscopic asteroseismological observing programs have also multiplied. While none have identified rotational splitting in an unevolved main sequence star, \cite{bouchy2005} observed rotational splitting in the acoustic spectrum of the G3IV-V star $\mu$ Arae, and several other groups have come close to an identification (see, e.g., \cite{bazot2007}). One of the most ambitious spectroscopic observing collaborations is the SONG project \citep{grundahl2008}, which aims to build several dedicated telescopes spaced in longitude around the world to allow for continuous observing. Spectroscopic observations should theoretically provide more precise stellar inclinations measurements than photometry, since the widths and shapes of the absorption lines provide additional information about the photosphere that is not present in luminosity variations. Work is currently underway to characterize the exact stellar inclination precision that can be expected from spectroscopic asteroseismology (T. Campante \& H. Kjeldsen, 2009, private communication). 

\subsection{Practical Applications}

Precise measurements of stellar inclinations are a step toward a practical transit survey of the brightest Sun-like stars in the sky. One can envision a space-based survey for transiting exoplanets targeted at these stars. The idea of manufacturing and launching a suite of nanosatellites, each with a single telescope and targeted at an individual star is currently under study \citep{seager2008}. The goal of the study would be to design and build a space telescope that fits within a 10 x 10 x 30 cm$^3$ triple CubeSat spacecraft, and to take advantage of the growing number of piggyback launch opportunities to place these telescopes into low-Earth-orbit. The present major challenge is the required pointing stability to achieve a high enough photometric precision to detect the transit of an Earth-analog on a low-mass satellite ($\leq 5$kg).

The duration of the survey would be largely set by the number of stars that would need to be observed photometrically for transits. Each star would need to be observed for at least a year to cover the full orbital period of an Earth-analog; though it may be possible to assign individual spacecraft multiple targets. For the initial stellar inclination measurements, using spectroscopic $v\sin(\psi)$ inclination determinations would require photometric observations spread over at least a stellar rotation period --- on the order of a month for solar-type stars. Asteroseismic inclinations would be, on average, faster to come by: SONG estimates that their network, with six dedicated observing stations across the globe, could measure the inclination of up to a few dozen stars over one year (H. Kjeldsen, 2009, private communication).

In terms of actual telescope time, the photometry for $v\sin(\psi)$ inclination determinations would require on average 10 minutes a night over the course of a month, for a total of 5 telescope-hours per star over that month. The asteroseismic inclination measurements, using SONG's estimates, would require about 10 telescope-days per star. Our fiducial Earth analog search would require inclination measurements on 830 stars, which corresponds to about 170 telescope-days for $v\sin(\psi)$ measurements, and about 8400 days for a 6 telescope SONG network. The photometric transit survey portion would take approximately 120 telescope-years. Note that even with the time required for the inclination measurements, this is one seventh the time required for a comparable blind survey.  

The idea of a targeted space-based transit search is made feasible by the enhanced transit probabilities that are achievable using stellar inclination constraints. One possible mission concept would be to use ground-based $v\sin(\psi)$ measurements to eliminate a third of the possible targets from consideration. Asteroseismology conducted from the ground and from orbiting triple CubeSat spacecraft could then be used to assemble a final target list for photometric observations. Depending upon the precision of the asteroseismology, the desired number of detections and the desired confidence level of achieving this many detections, this would reduce the number of stars that would need to be observed from a few thousand to a few hundred. A related proposal, using only ground-based asteroseismology has been described by \cite{beatty2009}.

\section{Summary}

Transit surveys of the brightest solar-like stars are made difficult by the distribution of these targets across the entire sky. This makes a traditional point-and-stare photometric survey unworkable, since each target star would need to be observed individually. Such a survey of the brightest (V $\leq7$) stars would require either a prohibitive number of telescopes or a prohibitive amount of observing time, especially if it is targeting Earth-analogs in stellar habitable zones. This problem motivated us to consider the effect of stellar inclination measurements on transit probabilities and the number of stars that need to be observed in a photometric survey to statistically expect a detection.

We derived the transit probability for stars, individually and in an ensemble, with constraints on the stellar inclinations and with assumptions about the range of planetary system inclinations. This derivation involved several steps. First, the stellar inclination constraint was treated as the conditional distribution in Bayes theorem and combined with our prior assumption of randomly oriented orbits. Second, the planetary inclination was included via a joint probability distribution together with the stellar inclination constraints. These steps completed the probability distribution for a single star. These calculations also allowed us to compute the transit probability for the second planet in a multiplanet system when the other planet has already been observed to transit. In this case, we used the orbital plane of the transiting planet in place of the stellar inclination as the reference plane for the system, and the mutual inclination of the two planets in place of the planetary system inclination. An ensemble of stars was treated using the binomial distribution and the distribution of expected transit detections to determine the number of stars required to be observed, and the probability that doing so would yield a detection. 

In summary, our first result is the transit probability of a single planetary system that has measured stellar inclination constraints. Assuming inclination measurements accurate to $5^\circ$, we find that the transit probabilities for typical hot Jupiters may be increased to 74.8\%, and the transit probabilities for Earth-analogs may be increased to 4.25\%. These calculations may also be applied directly to finding the transit probability of the second planet in a multiplanet system where one planet has already been observed to transit. For the specific case of HAT-P-13, we find that the transit probability for the outer planet is between 0-8.5\%, dependent on what is the assumed spread of the mutual inclination. Our second result is the estimated number of stars needed to be observed for transits given a specified required number of planet detections and desired probability of achieving those detections. Assuming $5^\circ$ uncertainty on the stellar inclination measurements, we would need to look at 120 stars to have a 95\% chance of detecting more than one Earth-analog. We would have a 50\% chance of detecting more than 3.8. This is one seventh the number of stars that a blind transit survey would need to look at for the same yield.

\acknowledgments 
We thank H. Kjeldsen and T. Campante for their helpful correspondence and useful calculations with regard to asteroseismology. We also thank Leslie Rogers for her comments and discussion. This work was supported in part by the NASA ASTID program.

\appendix

\section{Optimizing $P_{obs}$ and $n_\mathrm{i}$}

The overarching goal in optimizing the $P_\mathrm{obs}$ and $n_\mathrm{i}$ is to minimize the length of time needed to complete the survey. Recall that we start with the initial set of $n_\mathrm{i}$ target stars whose stellar inclinations we measure. The exact value of $n_\mathrm{i}$ will determine, through the statistics of the binomial distribution, the probability $P_\mathrm{ntr}$ that a given number of transiting planets are within our initial target stars. We then observe a fraction of these initial stars that have stellar inclinations close to $90^\circ$ photometrically for transits. The size of this fraction sets the probability $P_\mathrm{obs}$ that we observe photometrically the transiting planets hidden within our initial stars. As described in Equation (\ref{eq:5300}), the size of this fraction is dependent upon the precision of the stellar inclination measurements; the more precise the measurements, the smaller the fraction may be for a fixed $P_\mathrm{obs}$. 

Consider our notional survey to discover transiting Earth-analogs. As described in Section 4.3, it would take on average a month per star to measure stellar inclinations --- roughly enough time to see one complete rotation of a Sun-like star. The photometric transit observations would take at least one year per star. Assuming that every star has an Earth-analog, that stellar inclinations may be measured to $\pm 5^\circ$ and that planetary inclinations are evenly distributed within $7.5^\circ$ of the stellar equator, Figure 7 shows the effect of varying $P_{obs}$ and $n_\mathrm{i}$ on the time needed to complete the survey --- while keeping the overall detection probability constant at 0.95. As the number of initial targets increases, the corresponding fraction (and hence $P_\mathrm{obs}$) of those stars that need to be photometrically observed decreases. The minimum amount of time needed to complete the survey occurs when $n_\mathrm{i}=830$ and $P_\mathrm{obs}=0.7746$.  

\begin{figure}
\vskip -0.0in 
\epsscale{1.} 
\plotone{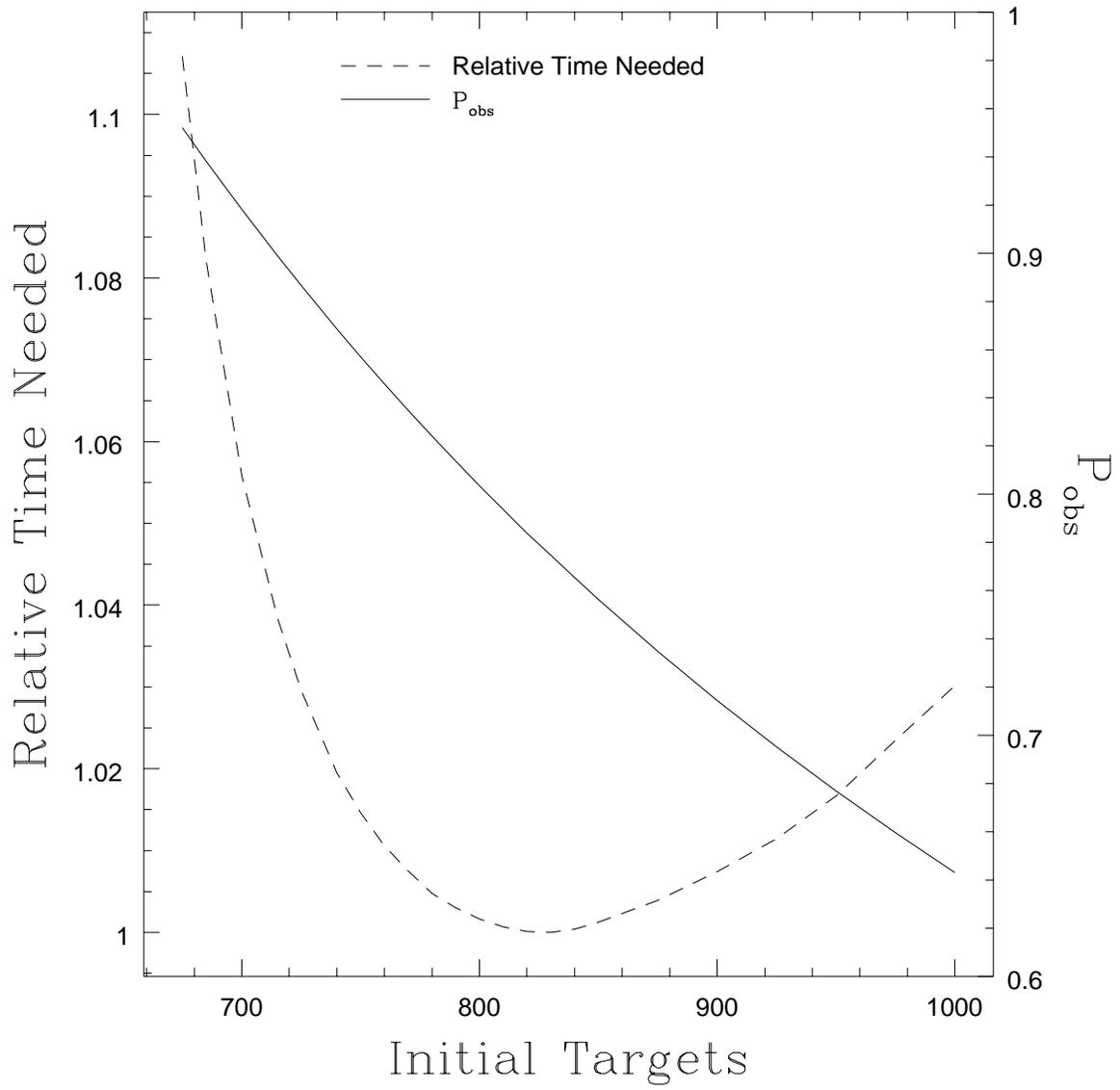}
\vskip -0.0in 
\figcaption{Optimization of the $P_{obs}$ and $n_{i}$ for an Earth-analog search that will have a 95\% confidence of at least one detection. The minimum amount of time required to complete the survey (normalized to unity in the figure) occurs when $P_\mathrm{obs}=0.775$ and $n_\mathrm{i}=830$.}
\end{figure}

While we do not show the optimization of a hot Jupiter survey in Figure 7, the method is exactly that same as for an Earth analog survey. The only major difference is that the photometric observations take about 10 days to complete. Since the time needed to measure the stellar inclination remains one month, this pushes a hot Jupiter survey to minimize the number of initial targets that must be observed for inclination measurements. If we wish a $95\%$ chance of detecting at least one hot Jupiter, then this means we will have to set $n_\mathrm{i}=29$ and $P_\mathrm{obs}=0.9814$ --- assuming that every star is orbited by a hot Jupiter.

\end{document}